\documentclass[aps,preprint]{revtex4}
\usepackage{amssymb}
\usepackage{amsmath}
\usepackage{graphicx}
\usepackage{epstopdf}
\begin{document}

\title{\bf Structure of Spin
Polarized Strange Quark Star in the Presence of Magnetic Field at Finite Temperature}

\author{{ G. H. Bordbar$^{1,2}$
\footnote{Corresponding author. E-mail:
bordbar@physics.susc.ac.ir}}, { F. Kayanikhoo $^{1}$} and { H. Bahri  $^{1}$}}
\affiliation{Department of Physics,
Shiraz University, Shiraz 71454, Iran\\
and\\ Research Institute for Astronomy and Astrophysics of Maragha,
P.O. Box 55134-441, Maragha 55177-36698, Iran}
\begin{abstract}
 In this paper, we have calculated the thermodynamic properties
of spin polarized strange quark matter at finite temperature
in the presence of a strong magnetic field using MIT bag model. We have also computed
the equation of state of spin polarized strange quark matter in the presence
of strong magnetic field and finally, using this equation of states we have investigated
 the structure of spin
polarized strange quark star at different temperatures and magnetic fields.
\end{abstract}
\maketitle

 %%%%%%%%%%%%%%%%%%%%%%%%%%%%%%%%%%%%%%%%%%%%%%%%%%%%%%%%%%%%%%%%%%%%%%%%%%%%%%%%%%%%%
\section{Introduction}
One of the interesting  features of the compact objects is the
strange quark star (SQS) composed of the strange quark matter (SQM).
The composition of SQS was first proposed by Itoh \cite{rk3}; simultaneously
with formulation of Quantum chromo dynamics (QCD).
Later, Bodmer discussed the fate of an astronomical object collapsing
to such a state of matter \cite{rk4}. The perturbative computations of the equation of state of the SQM
was developed after the formation of QCD, but the region
of validity of these calculations was restricted to very high densities \cite{rk50}.
The concept of SQS was also discussed by witten. He proposed that SQM composed of light quarks, and is more stable
than the nuclei \cite{rk5}. Therefore the SQM can be the ground state of matter.
With this point of view, other authors proposed the concept of SQS.
The SQM is composed of 3-flavors of quarks (up, down and strange)
and a small number of electrons to ensure the charge neutrality
\cite{rk6,rk6p}. A typical electron fraction
is less than $10^{-3}$ $(fm^{-3})$ and it decreases from the surface to the center of SQS.

The collapsing star of a supernova will turn into a neutron star only if its mass is about
$1.4-3$ $M_{\odot}$. After formation of the neutron star, if density of core is high enough ($10^{15} \ gr/cm^{3}$)
the nucleons dissolve to their components, quarks, and a hybrid star (neutron star with a core of SQM) is formed .
But if after the explosion of supernova, in the stage of proto-neutron,
the density of the core is high enough ($10^{15} \ gr/cm^{3}$), the pure SQS may be formed, directly. Making a quark-novae.
If after collapse the leftover is more than $3M_{\odot}$ mass the star will be transform to a black hole.
For a neutron star, the radius decrease with mass, or in other word $ M \propto \ 1/R$; while for an SQS,
the mass-radius relation is as $M \propto \ R^{3}$. Some result of observations show that for  a typical SQS, the mass
is more than $ 2.1\pm 0.28 \ M_{\odot} $ and the radius is more than $ 13.8\pm1.8 \ km $ \cite{rk7}.
Besides, recent observations indicate that objects RX J185635-3754 and 3C58 may be SQS \cite{rk51}.

The maximum magnetic field on the surface of magnetars is about $10^{15} \ G$,
however the highest possible magnetic field in the center of SQS that
was estimated by virial theorem is between $10^{18} - 10^{20} \ G$ \cite{rk75,rk62}.
 Therefore, investigating the effect of an strong
 magnetic field on the SQM properties is important in astrophysics.
 Some studies that investigate the effect of magnetic field on quark star and quark matter
 are available.  Chakrabarty \cite{rk63}
 studied the effects of an strong magnetic field on quark matter.
 He showed that the equation of state of SQS changes significantly at the presence of strong magnetic fields.
 Anand et.al. \cite{rk64} expressed that the maximum mass,
 the radius and gravitational red-shift of the SQS are increasing as a function of the magnetic field.

Recently, we have calculated the structure of polarized SQS at zero temperature \cite{rk8}
and unpolarized SQS at finite temperature \cite{rk9}.
We have also calculated the structure of the neutron star with the quark core at zero temperature
\cite{rk10} and at finite temperature \cite{rk11}.
In this paper we calculate some properties
of polarized SQS at finite temperature in the presence of an strong magnetic field.
The plan of this work is as follows.
In section \ref{II}, we calculate the energy and equation of state of SQM by the MIT bag model
at finite temperature in the presence of a strong magnetic field and in section \ref{sec4},
we present the results of our calculations for different
temperatures and magnetic fields.

\section{Calculation of energy and equation of state of strange quark matter}
\label{II}
The equation of state plays an important role in obtaining the structure of a star.
There are different statistical approaches for calculation of the equation of state of quark matter (SQM)
and all of them are based on the QCD, for example the MIT bag model
\cite{rk20,rk21}, NJL model \cite{rk22,rk23} and perturbation QCD model \cite{rk24,rk25}.
In this paper we calculate the equation of state of SQM by MIT bag model with a density dependent bag constant.
The MIT bag model considers free particles confined to a bounded region by a bag pressure $B$.
This pressure depends on the quark-quark interaction \cite{rk55}.
In other words, the bag constant ($B$) is the difference between the energy densities of the
noninteracting quarks and the interacting ones.
For the bag constant in the MIT bag model, different values such as $55$ and $90 \ MeV/fm^{3}$  are considered.

To obtain the equation of state of SQM, at first, we should calculate the total energy of the system.
We do this work using the MIT bag model with a density dependent bag constant in the next section.

\subsection{Energy of spin polarized strange quark matter at finite temperature in the presence of magnetic field}
At densities high enough ($10^{15} \ gr/cm^{3}$) the hadronic phase is allowed to undergo a phase transition
to the strange quark matter phase. Strange quark matter consists of $u$, $d$ and $s$ quarks, as well as electrons in weak equilibrium;
\begin{equation}\label{01}
d \rightarrow u+e+\overline{\vartheta}_e,
\end{equation}
\begin{equation}\label{02}
u+e \rightarrow d+\vartheta_e,
\end{equation}
\begin{equation}\label{03}
u+d \rightarrow u+s,
\end{equation}
\begin{equation}\label{04}
s \rightarrow u+e+\overline{\vartheta}_e.
\end{equation}
To calculate the energy of SQM, we need to know the quark densities in term of the baryonic number density.
To compute these densities, we use the beta equilibrium and charge neutrality conditions in the above reactions.
The beta equilibrium and charge neutrality conditions lead to the following relations for the chemical potentials and densities of relevant quarks,
\begin{equation}\label{1}
\mu_{d} = \mu_{u}+\mu_{e},
\end{equation}
\begin{equation}\label{2}
\mu_{s} = \mu_{u}+\mu_{e},
\end{equation}
 \begin{equation}\label{3}
\mu_{s} = \mu_{d},
\end{equation}
\begin{equation}\label{4}
\frac{2}{3} n_{u}-\frac{1}{3} n_{s}-\frac{1}{3} n_{d}-n_{e} = 0,
\end{equation}
where $\mu_{i}$ and $n_{i}$ are the chemical potential and the number density of particle $i$ respectively. It should be noted that because the neutrinos escape from the interior matter of star, the chemical potential of neutrinos vanishes ($\mu_{\vartheta_e}=\mu_{\overline{\vartheta}_e}=0$). Therefore, we get the above relations between the chemical potentials \cite{rk6,rk6p}. As it was mentioned in the previous section,
we consider the system as pure SQM  and ignore the electrons($n_{e}=0$) \cite{rk13,rk14,rk15}, therefore Eq. (\ref{4}) becomes as follows,
\begin{equation}\label{5}
n_{u}=\frac{1}{2}(n_{s}+n_{d}).
\end{equation}

Now, we want to calculate the energy of SQM in the presence of an strong magnetic field. Therefore,
we consider the spin polarized SQM composed of the $u$, $d$ and $s$ quarks with up and down spins.
We show the number density of spin-up quarks by $n_{i}^{+}$ and spin-down quarks by $n_{i}^{-}$.
The parameter of polarization is defined by the following relation,
\begin{equation}\label{6}
\zeta_{i}=\frac{n_{i}^{+}-n_{i}^{-}}{n_{i}},
\end{equation}
where $0\leq \zeta_{i}\leq 1$ and $n_{i}=n_{i}^{+}+n_{i}^{-}$.
The chemical potential $\mu_{i}$ for any value of the temperature ($T$) and number density ($n_{i}$), using the following constraint,
\begin{equation}\label{7}
n_{i}=\sum_{P=\pm}\frac{g}{2\pi^{2}}\int_{0}^{\infty} f(n_{i},k^{(p)},T)k^{2}dk,
\end{equation}
where
\begin{equation}\label{8}
f(n_{i},k^{(p)},T)=\sum_{P=\pm}\frac{1}{exp\left(\beta((m_{i}^{2}c^{4}+
\hbar^{2}(k^{(p)})^2c^{2})^{1/2}-\mu_{eff})\right)+1},
\end{equation}
is the Fermi-Dirac distribution function. In the above equation $\beta=1/k_{B}T$ and $g$ is degeneracy number of the system. In this equation $\mu_{eff}$ is the effective chemical potential in the presence of magnetic field. The effective chemical potential is equal to $\mu_i \mp \mu_s B$ where $\mu_s$ is magnetic moment and $B$ is the magnetic field. In other word, the single particle energy is
$\varepsilon_i = \sqrt{m_{i}^{2}c^{4}+ \hbar^{2}k^p} \pm \mu_s B$.

The energy of spin polarized SQM in the presence of the magnetic field within the MIT bag model is as follows,
 \begin{equation}\label{9}
\varepsilon_{tot}=\varepsilon_{u}+\varepsilon_{d}+\varepsilon_{s}+\varepsilon_{M}+B,
\end{equation}
where
\begin{equation}\label{10}
\varepsilon_{i}=\sum_{P=\pm}\frac{g}{2\pi^{2}}\int_{0}^{\infty}(m_{i}^{2}c^{4}+
\hbar^{2}(k^{(P)})^{2}c^{2})^{1/2}f(n_{i},k^{(p)},T)k^{2}dk,
\end{equation}
here $k^{\pm}=(\pi^{2}n_{i})^{1/3}(1\pm\zeta_{i})^{1/3}$.  In our calculations, we suppose
that $\zeta=\zeta_{u}=\zeta_{d}=\zeta_{s}$. In Eq. (\ref{9}), $\varepsilon_{M}=\frac{E_{M}}{V}$  is the magnetic energy
density of SQM where $E_{M}=-M.B$ is the magnetic energy.
By considering a uniform magnetic field along $z$ direction, the contribution of magnetic energy of
the spin polarized SQM is given by
\begin{equation}\label{11}
E_{M}=-\sum_{i=u,d,s}M_{z}^{(i)}B,
\end{equation}
where $M_{z}^{(i)}$ is the magnetization of the system corresponding to particle $i$ which is given by
\begin{equation}\label{12}
M_{z}^{(i)}=N_{i}\mu_{i}\zeta_{i}.
\end{equation}
In the above equation $N_{i}$ and $\mu_{i}$ are the number density and magnetic moment of particle $i$.
Finally, the magnetic energy density of SQM can be obtained by the following relation,
\begin{equation}\label{13}
\varepsilon_{M}=-\sum_{i}n_{i}\mu_{i}\zeta_{i}B.
\end{equation}

We obtain the thermodynamic properties of the system using the Helmholtz free energy,
\begin{equation}\label{16}
F=\varepsilon_{tot}-TS_{tot},
\end{equation}
where $S_{tot}$ is the total entropy of SQM,
\begin{equation}\label{14}
S_{tot}=S_{u}+S_{d}+S_{s}.
\end{equation}
In Eq. (\ref{14}), $S_{i}$ is entropy of particle $i$,
\begin{eqnarray}\label{15}
S_{i}(n_{i},T) &=&-\frac{3}{\pi^{2}}k_{B}\int_{0}^{\infty}f(n_{i},k,T)ln(f(n_{i},k,T))\\ \nonumber
&+& (1-f(n_{i},k,T))ln(1-f(n_{i},k,T))]k^{2}dk.
\end{eqnarray}

\subsection{Equation of state of spin polarized strange quark matter}
For deriving the equation of state of strange quark matter (SQM), the following equation is used,
\begin{equation}\label{17}
P(n,T)=\sum_{i}(n_{i}\frac{\partial F_{i}}{\partial n_{i}}-F_{i}),
\end{equation}
where $P$ is the pressure of system, $F_i$ is the free energy of
particle $i$ and $n_i$ is the numerical density of particle $i$.

%---------------------------------------------------------------------------------------------

\section{Results and discussion}
\label{sec4}

\subsection{Thermodynamic properties of spin polarized strange quark matter}

\begin{figure}
\centering
\includegraphics[scale=0.45]{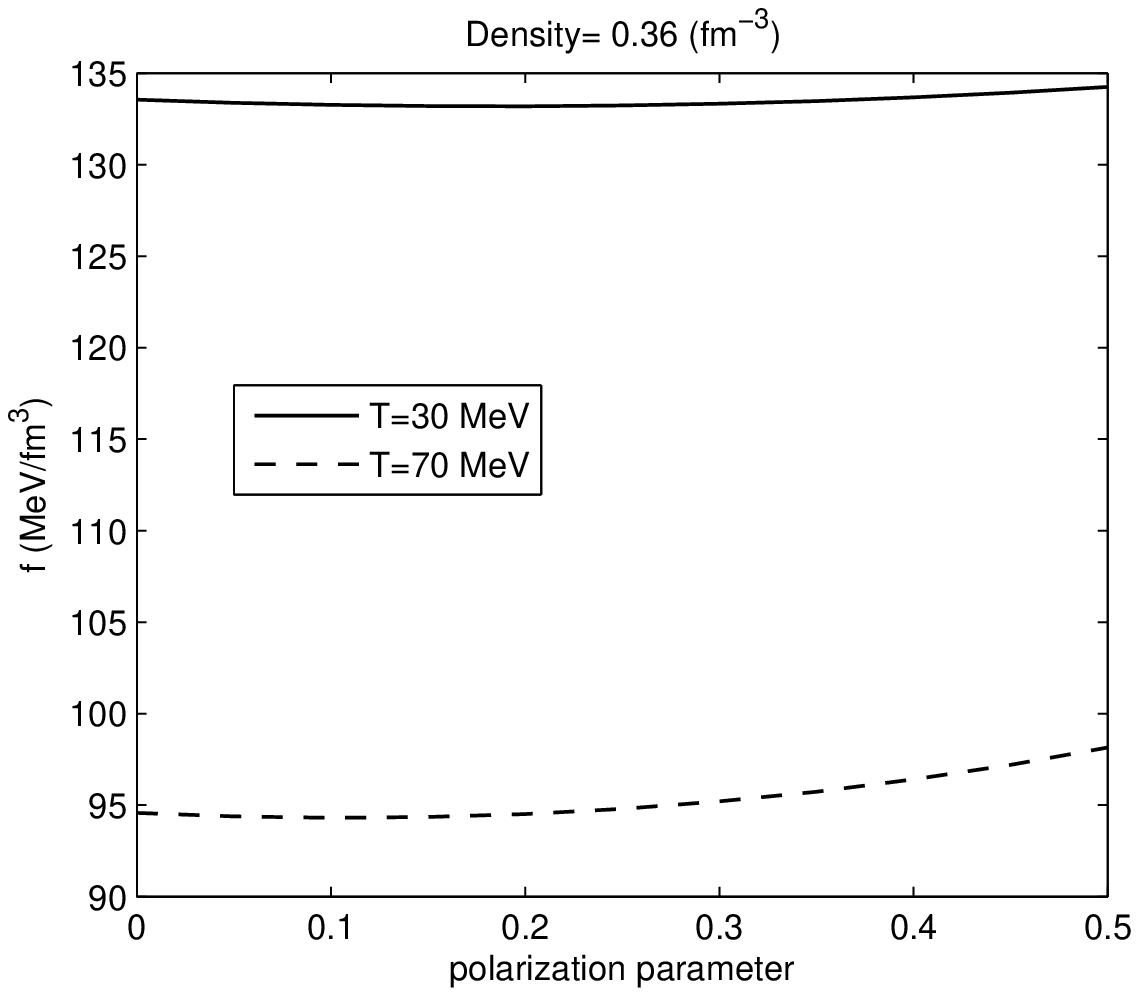}\quad
\includegraphics[scale=0.45]{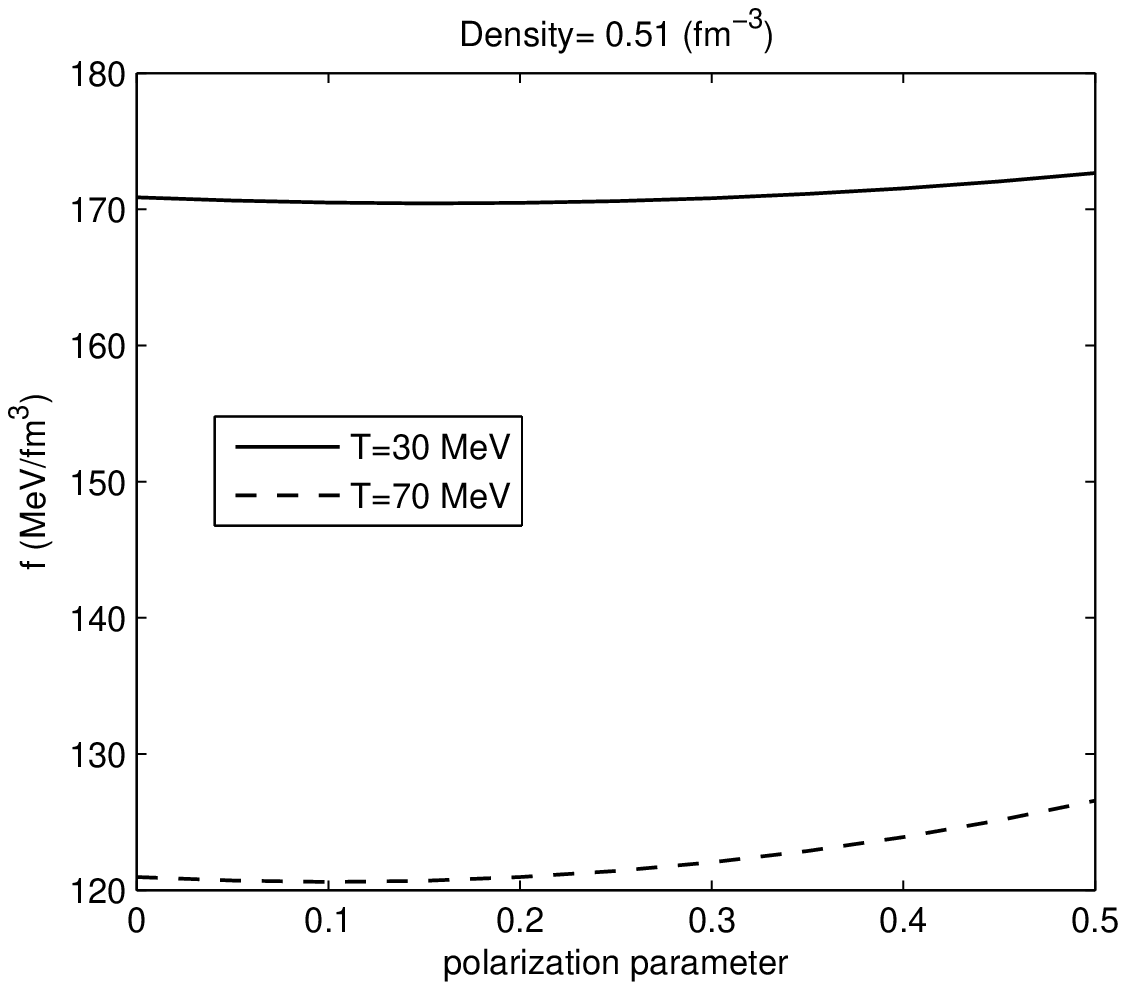}\quad
\includegraphics[scale=0.45]{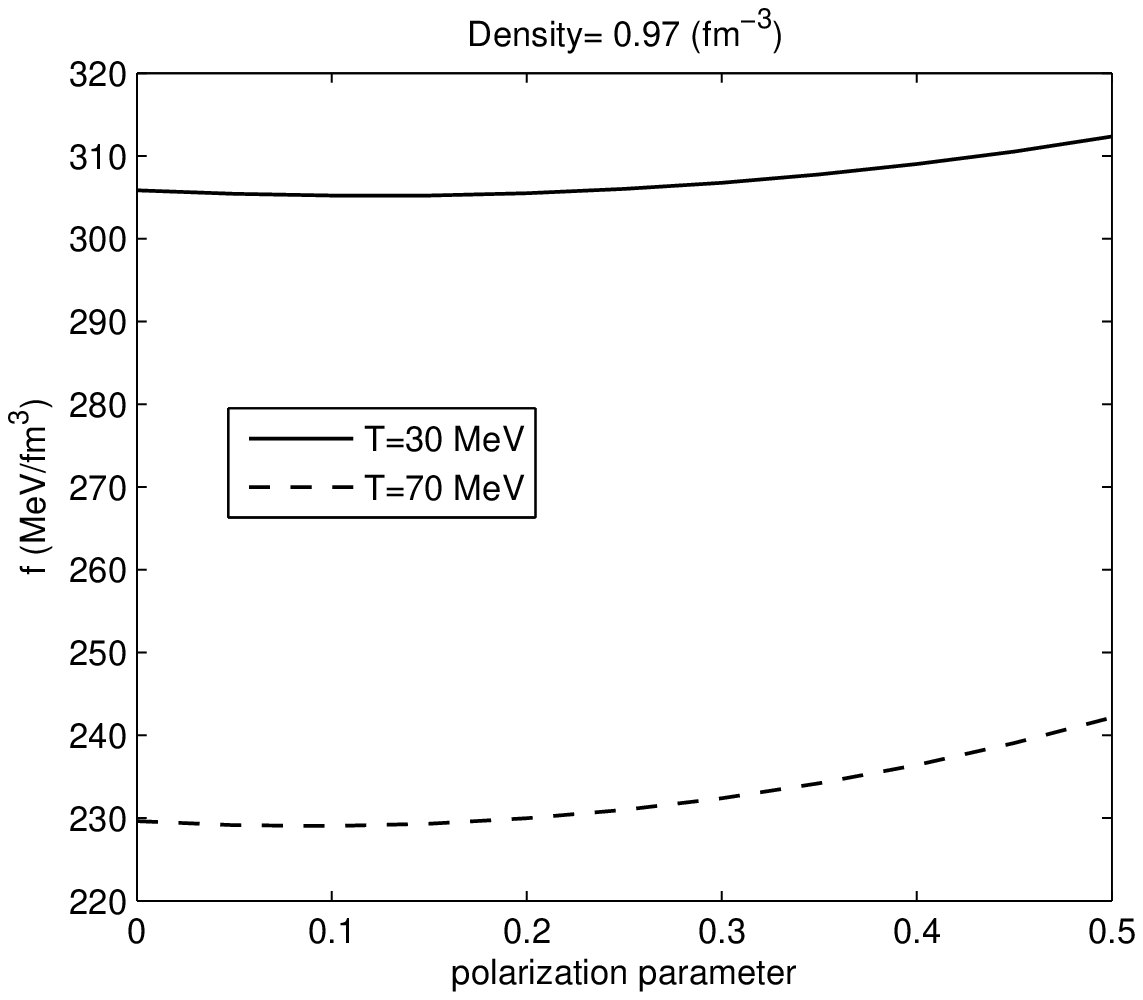}\quad
\includegraphics[scale=0.45]{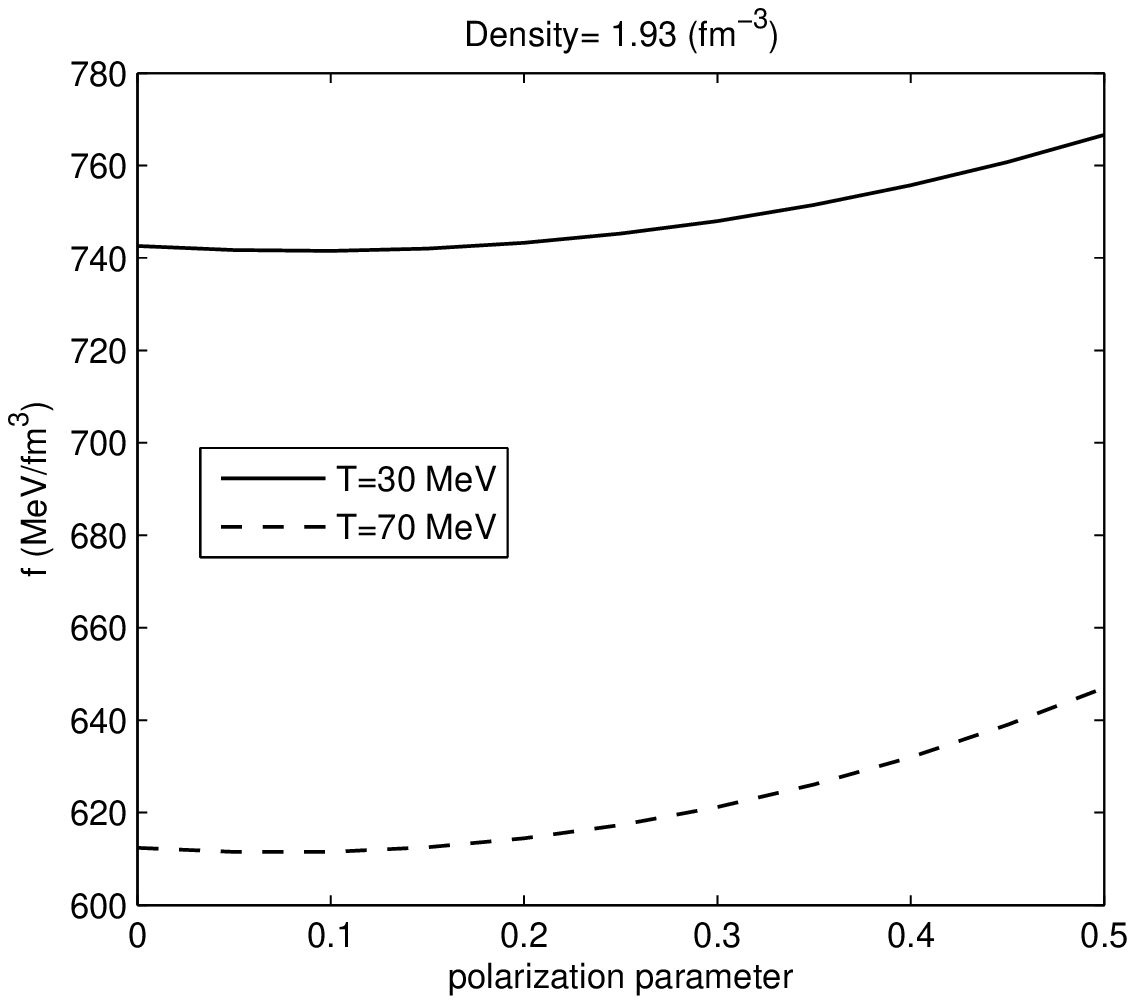}

\caption{Total free energy per volume as a function of the polarization parameter at different
temperatures and densities for $B=5\times10^{18}\ G$.} \label{fig1}
\end{figure}

In Fig. \ref{fig1}, we present the total free energy per volume
versus the polarization parameter $\zeta$, for $B=5\times10^{18} \ G$,
at different densities and temperatures. From Fig. \ref{fig1}, we
can see that at each density and temperature, the energy is
minimum at a particular value of the polarization parameter. This
indicates that at each density and temperature, there is a meta
stable state. We have found that for a fixed temperature, the
polarization parameter corresponding to the meta stable state
reaches zero by increasing the density. Similarly, we can see
that at a fixed density, in the meta stable state the polarization parameter
decreases by increasing the temperature. These results agree with
those of reference \cite{rk94}.

\begin{figure}
\centering
\includegraphics[width=\textwidth, angle=0]{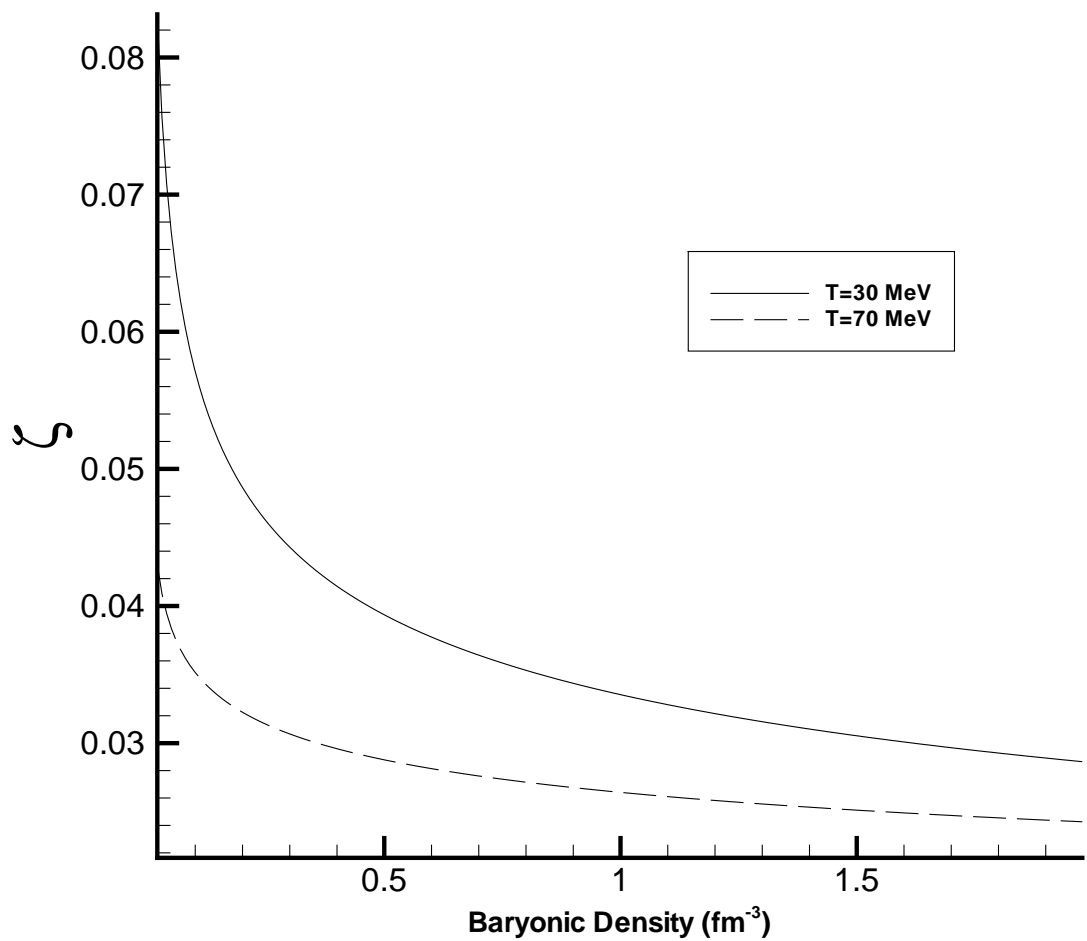}
\caption{The polarization parameter versus baryonic density at different temperatures for $B=5\times10^{18}\ G$.} \label{fig2}
\end{figure}

\begin{figure}
\centering
\includegraphics[width=\textwidth, angle=0]{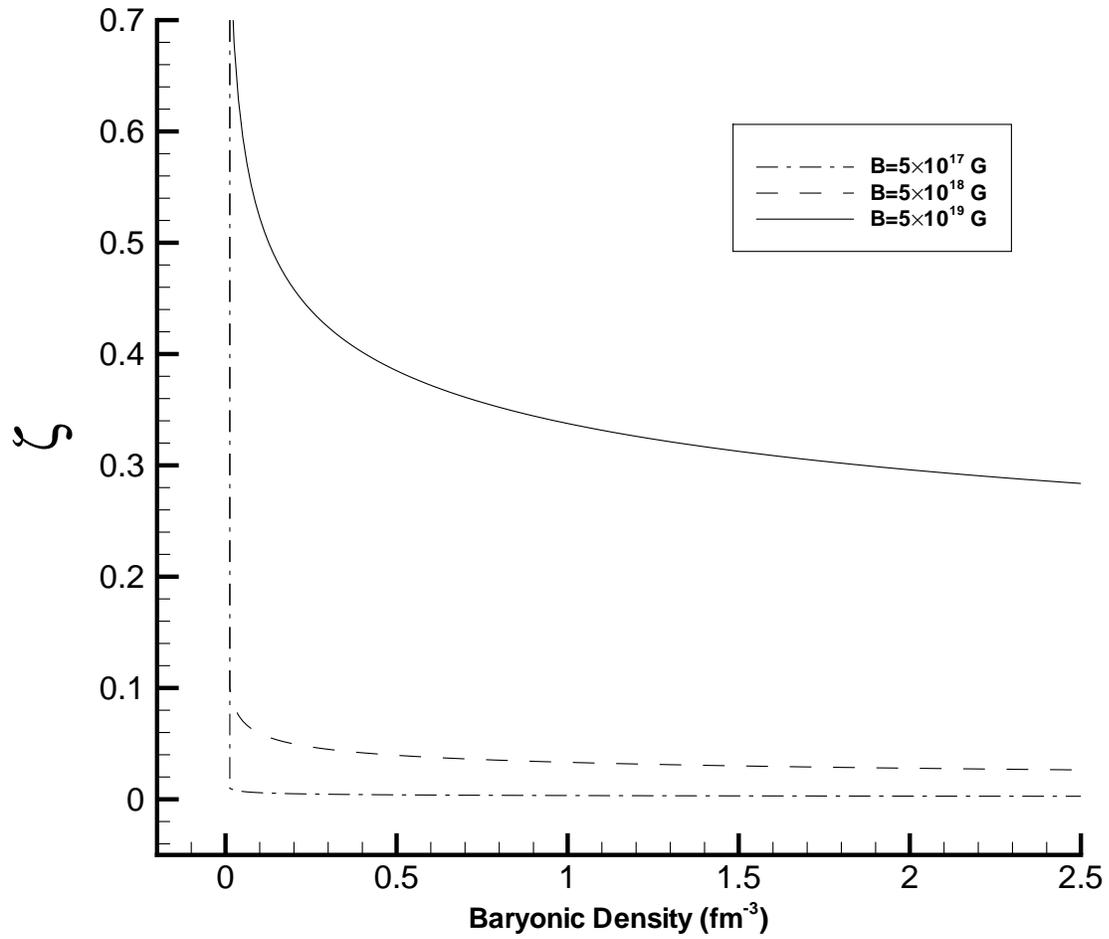}
\caption{The polarization parameter versus baryonic density at
$T=30\ MeV$ for
different magnetic fields.} \label{fig3}
\end{figure}

\begin{figure}
\centering{
\includegraphics[width=\textwidth, angle=0]{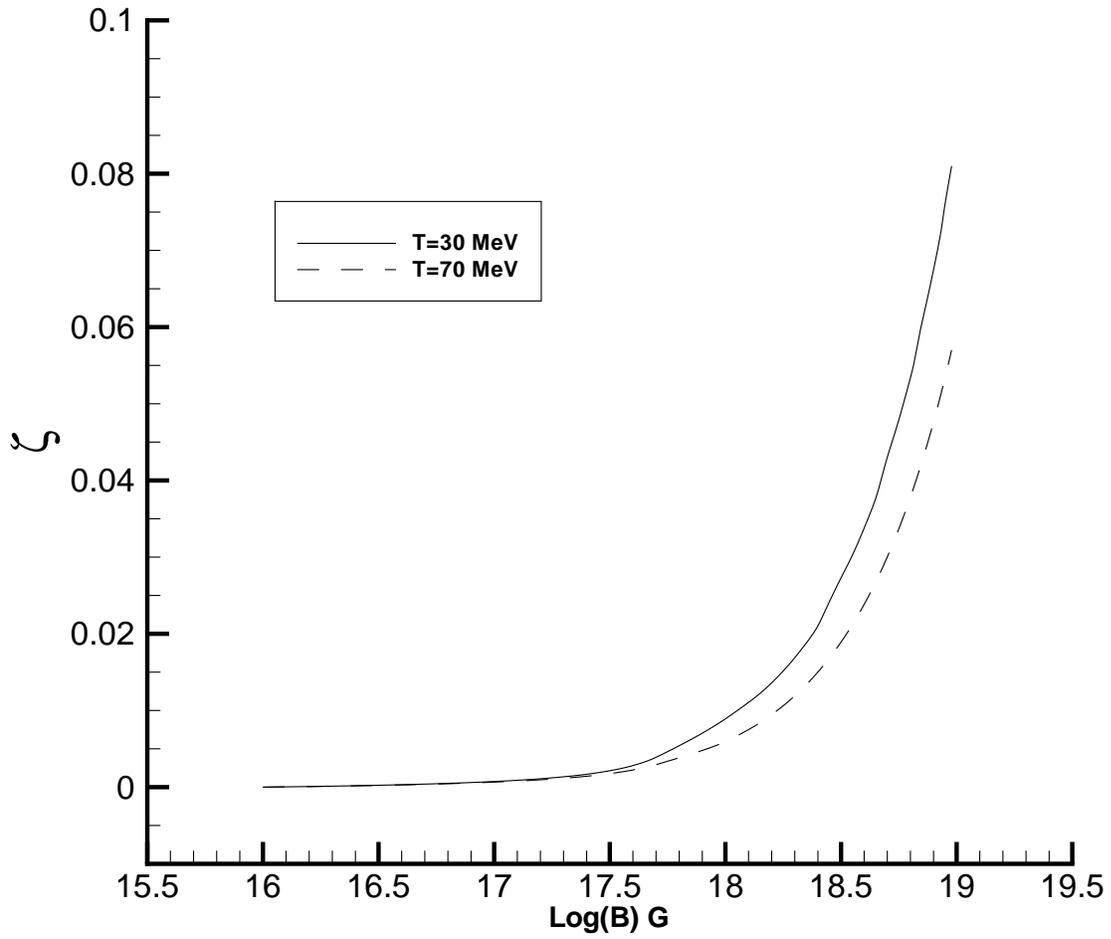}
\caption{The polarization parameter as a function of magnetic field at
$n=0.5\ fm^{-3}$ for different temperatures.} \label{fig4}}
\end{figure}

In Fig. \ref{fig2}, we have plotted the polarization parameter versus
baryonic density at different temperatures in the presence of
magnetic field $B=5\times10^{18} \ G$. We can see  that
the polarization parameter decreases by increasing the baryonic density.
Fig. \ref{fig2} shows that at a fixed density, the polarization parameter decreases  by
increasing the temperature.
This is due to the fact that contribution of magnetic energy is more significant
at low temperature.
The polarization parameter has been also drawn versus baryonic
density at a fixed temperature ($T=30 \ MeV$) for different magnetic fields in Fig. \ref{fig3}.
It is seen that for each density,  the polarization parameter increases by increasing
the magnetic field. As it can be seen from  Fig.\ref{fig3},
at high baryonic densities, for the magnetic fields lower than about $5\times10^{18} \ G$
the polarization parameter nearly becomes zero.
In other words,  for lower magnetic fields at higher densities, the strange quark matter
(SQM) becomes nearly unpolarized.
In Fig. \ref{fig4}, we have shown the polarization parameter as a function of the magnetic field for
density $n=0.5 \ fm^{-3}$ at different temperatures.
We can see that at lower values of the magnetic field, the polarization parameter
is nearly zero, and at the magnetic fields greater than about $5\times10^{17} \ G$,
the polarization parameter increases as the magnetic field increases.
Also, this figure shows that for each value of the magnetic field, the polarization parameter increases
by decreasing the temperature.
This is because at a higher temperatures the contribution of the kinetic energy
is higher compared to that of the magnetic energy.

\begin{figure}
\centering
\includegraphics[width=\textwidth, angle=0]{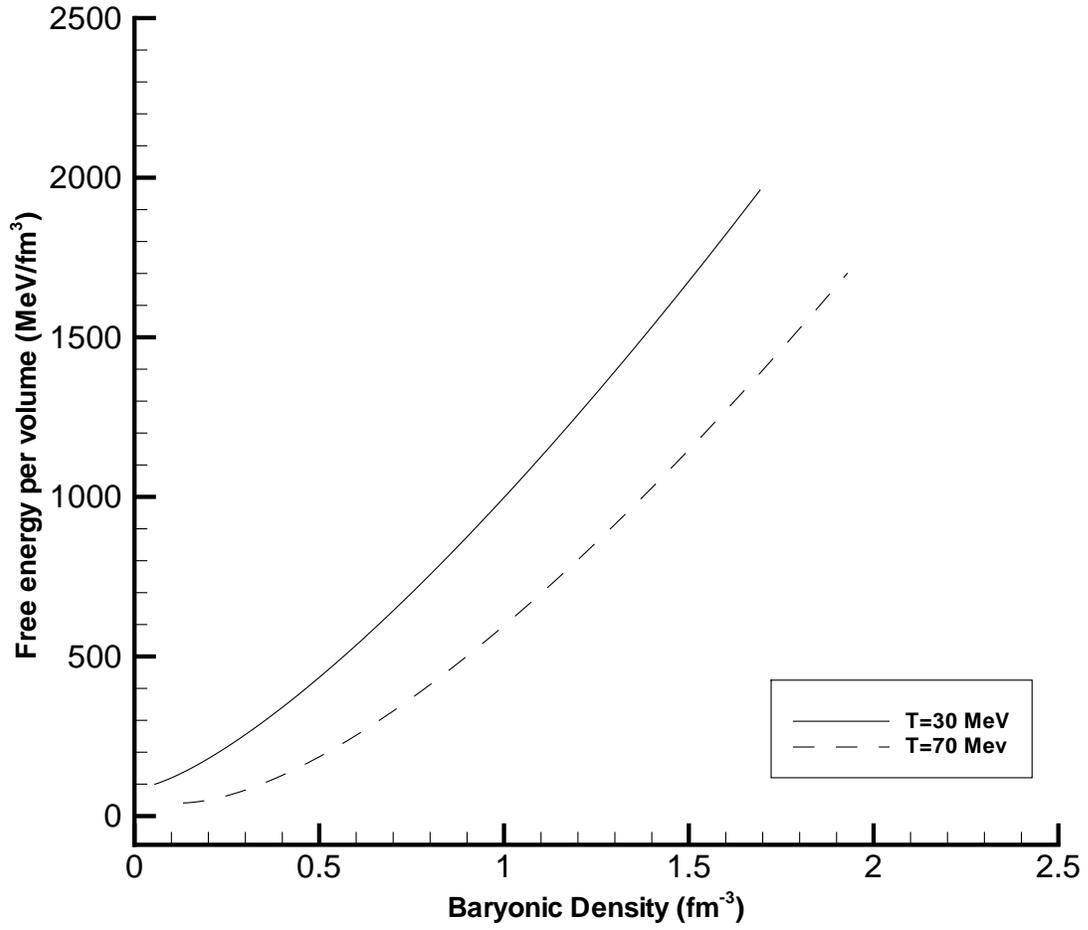}
\caption{The total free energy per volume of the spin polarized SQM as a function of the baryonic
density for $B=5\times10^{18}\ G$ at different temperatures.} \label{fig5}
\end{figure}

We have shown the total free energy per volume of the spin polarized SQM
as a function of the baryonic density in Fig. \ref{fig5} for the magnetic
field $B=5\times10^{18} \ G$ at different temperatures.
From this figure,we have found that the free energy of the spin polarization
SQM has positive values for all temperatures and densities.
For each density, it is shown that the free energy decreases by increasing the temperature.

\begin{figure}
\centering
\includegraphics[width=\textwidth, angle=0]{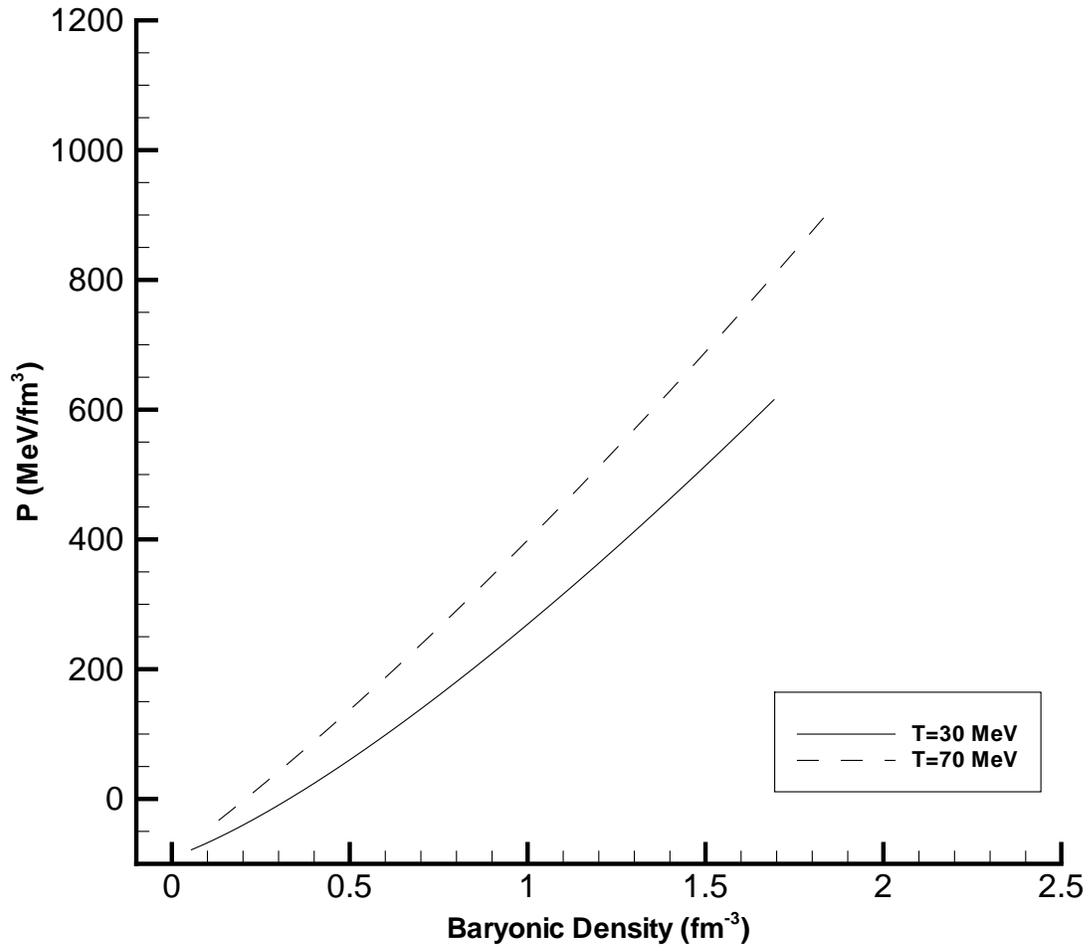}
\caption{The pressure of the spin polarized SQM for $B=5\times10^{18}\ G$  at different temperatures.} \label{fig6}
\end{figure}

The pressure of spin polarized SQM at different temperatures in the presence of
magnetic field ($B=5\times10^{18} \ G$) has been plotted in Fig. \ref{fig6}.
This figure shows that the pressure of SQM increases
by increasing the density. We can also see that  the pressure increases by increasing temperature.
These results indicate that the equation of state
of spin polarized SQM becomes stiffer by increasing the temperature.

\subsection{Structure of spin polarized strange quark star}
\label{struc}
The structures of compact objects such as white dwarfs,
neutron stars, pulsars and strange quark stars (SQS), are determined
from the Tolman-Oppenheimer-Volkov equations (TOV) \cite{rk16},
\begin{equation}\label{18}
\frac{dP}{dr}=-\frac{G\left[\varepsilon(r)+\frac{P(r)}{c^{2}}\right]\left[m(r)+\frac{4\pi
r^{3}P(r)}
{c^{2}}\right]}{r^{2}\left[1-\frac{2Gm(r)}{rc^{2}}\right]},
\end{equation}
\begin{equation}\label{19}
\frac{dm}{dr}=4\pi r^{2}\varepsilon(r).
\end{equation}
Using the equation of state which was obtained in the previous
section, we integrate the TOV equation to compute the structure of
SQS.

\begin{figure}
\centering
\includegraphics[width=\textwidth, angle=0]{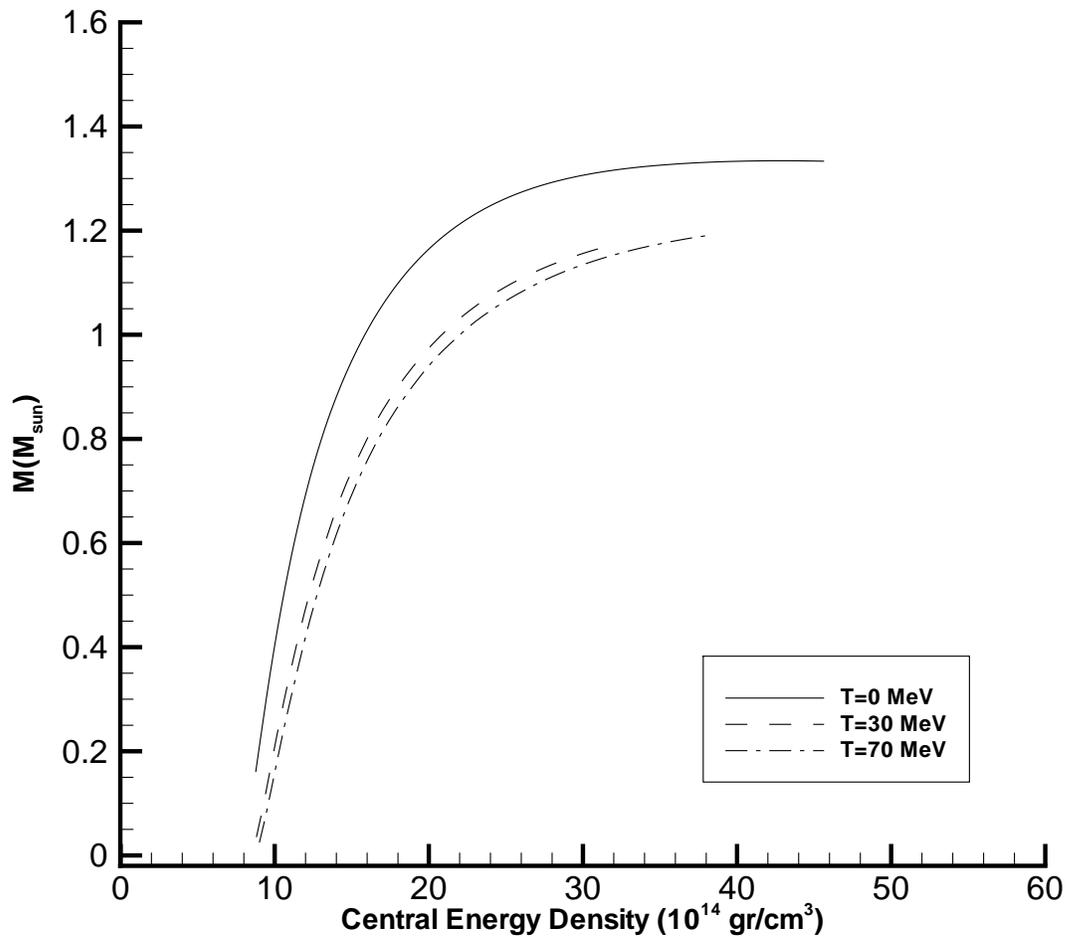}
\caption{The gravitational mass of spin polarized SQS versus energy density at different temperatures
for $B=5\times10^{18}\ G$. The results of  $T=0\ MeV$ (Bordbar \& Peyvand~\cite{rk8}) have been
  also given for comparison.} \label{fig7}
\end{figure}

In Fig. \ref{fig7}, we have plotted the gravitational mass of spin polarized strange quark star
(SQS) versus energy density at different temperatures for a magnetic field $B=5\times10^{18} \ G$.
We can see that for all temperatures, the gravitational mass
increases rapidly by increasing the energy
density and finally reaches a limiting value (maximum gravitational mass).
Fig. \ref{fig7} shows that this limiting value decreases by increasing the temperature.
From this result, we can conclude that the stiffer equation of state leads to the lower
values for the gravitational mass of SQS.

\begin{figure}
\centering
\includegraphics[width=\textwidth, angle=0]{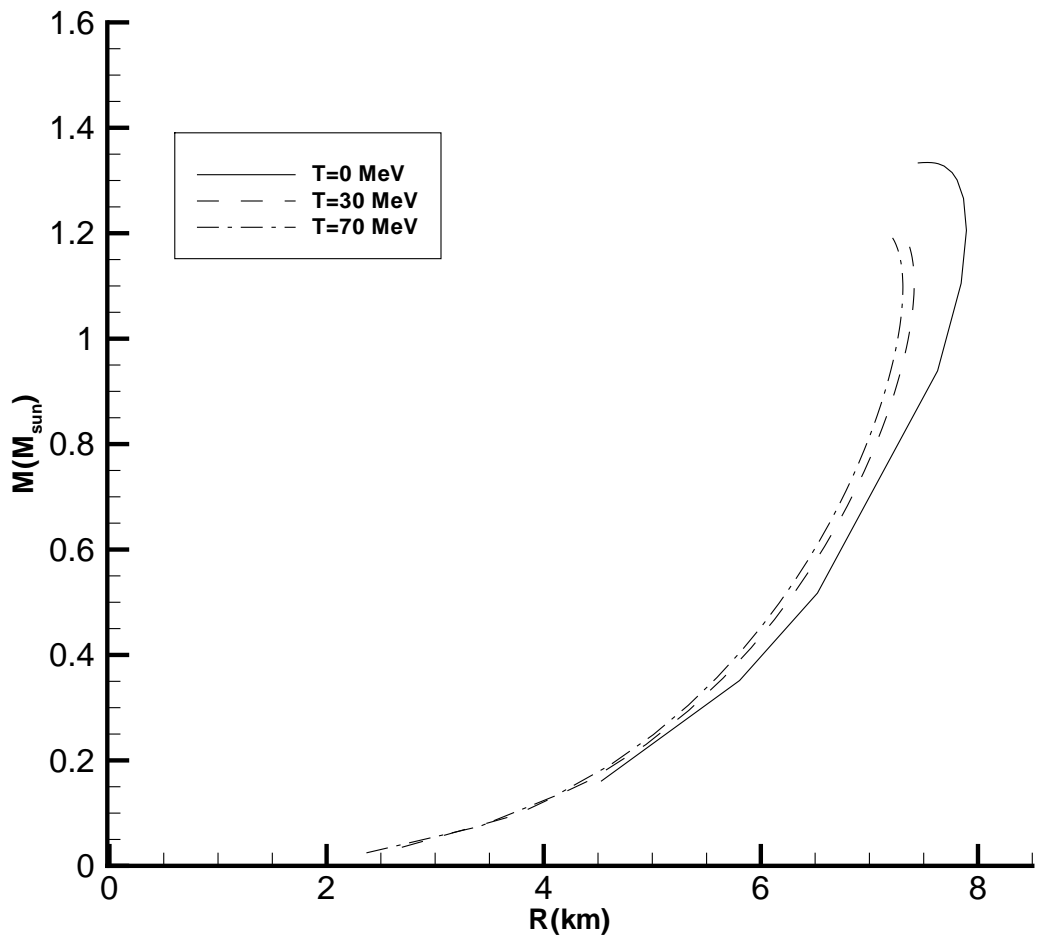}
\caption{The gravitational mass of spin polarized SQS as a function of the radius at different temperatures
for $B=5\times10^{18}\ G$. The results of  $T=0\ MeV$ (Bordbar \& Peyvand~\cite{rk8}) have been
  also given for comparison.} \label{fig8}
\end{figure}

\begin{figure}
\centering
\includegraphics[width=\textwidth, angle=0]{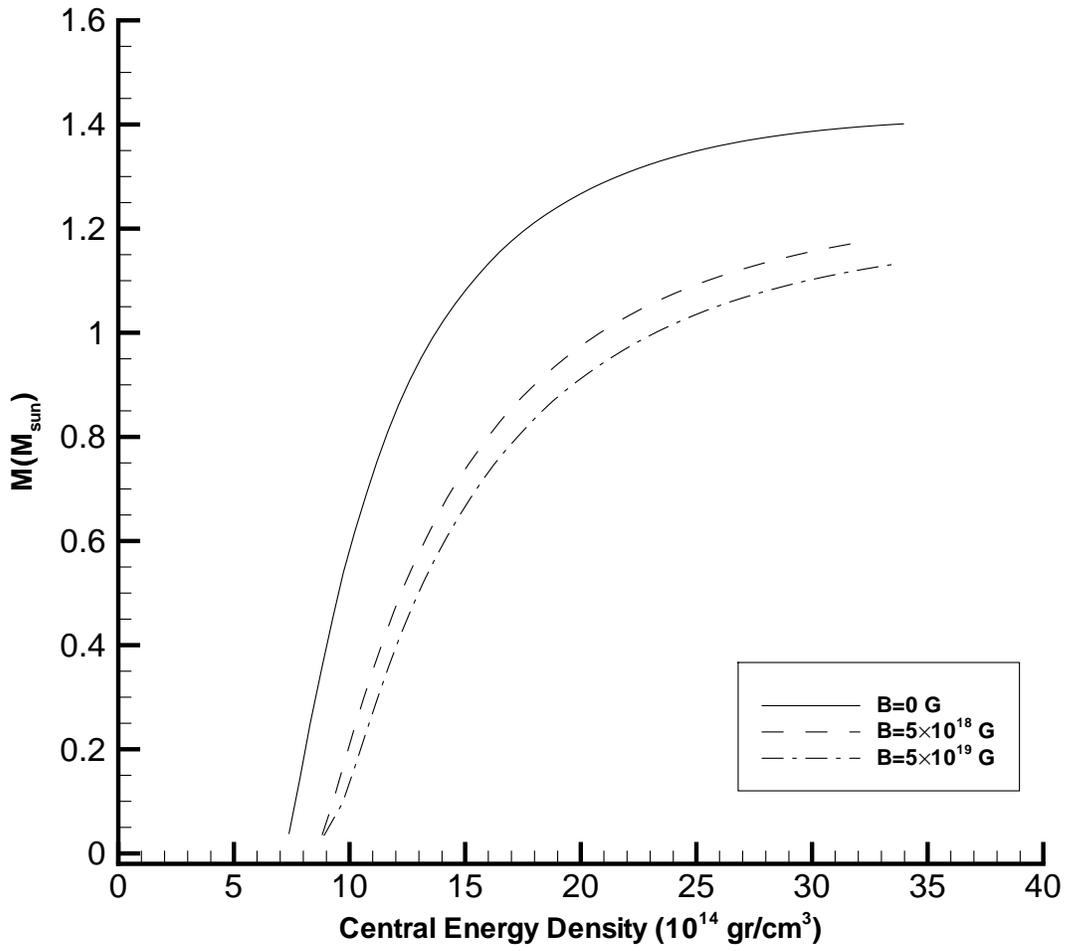}
\caption{The gravitational mass versus energy density at  $T=30\ MeV$ for different magnetic fields.} \label{fig9}
\end{figure}

We have shown the gravitational mass of spin polarized SQS as a function of the radius at different
temperatures in Fig. \ref{fig8}.
This figure shows that the gravitational mass  increases  by increasing radius.
We can see that the increasing rate of gravitational mass versus radius
increases by increasing the temperature.
According to Fig. \ref{fig6}, by increasing temperature, the equation of state becomes stiffer
leading to the conclusion that by increasing temperature,
mass and the radius of spin polarized SQS decrease.
%
%%%%%%%%
Fig. \ref{fig9} shows the gravitational mass of SQS as a function of the energy density at a fixed
temperature ($T=30 \ MeV$) for different magnetic fields. It can be seen that as the magnetic field increases,
the rate of increasing of gravitational mass versus central energy density decreases,
getting to a lower value at the limiting case.
%%%%%%%
In Table \ref{T1}, we have shown the maximum mass and the corresponding radius of an spin polarized SQS at
different temperatures for $B=5\times10^{18} \ G$.
It is shown that by increasing the temperature,
the maximum mass and radius of spin polarized SQS decreases.
We have also shown the maximum mass and the corresponding radius of the spin polarized
SQS for different magnetic fields at a fixed temperature $T=30 \ MeV$ in Table \ref{T2}.
We see that the maximum mass and radius of the spin polarized SQS
decreases by increasing the magnetic field.
One of the method to compare the mass and radius in our calculation with the observational values is
calculation of the surface red-shift value.
The maximum surface red-shift calculated in our work is $z_s = 0.38$
(for $T=30 \ MeV$ $B = 5\times10^{19} \ G$ ) that is $55.34 $ lower than the upper bound on
the surface red-shift for sub-luminal equation of states (i.e. $z_s^{CL} = 0.8509$).
Furthermore the minimum amount of red-shift in our calculations is equal to $z_s = 0.37$ that is
almost near the red-shift of RX J185635-3754 is $z_s = 0.35\pm =0.37$.
%================================================================================================
\section{Summary and conclusions}
In this paper, we investigated the thermodynamic properties of the strange quark matter (SQM)
composed of the spin-up and spin-down of the up, down and strange quarks at finite temperature in the
presence of an strong magnetic field  using the MIT bag model.
We computed the total free energy of the spin polarized SQM at finite
temperature in the presence of an strong magnetic field. We found that the free energy gets a
minimum at a particular value of the polarization parameter showing
a meta-stable state.
We have also shown that the polarization parameter corresponding
to the meta-stable state reaches zero by increasing both density and temperature.
In other words, by increasing the density and
the temperature, the spin polarized SQM becomes nearly unpolarized.
We calculated the equation of state of the spin polarized SQM at different
temperatures for $B = 5\times10^{18} \ G$. We showed that
by increasing temperature, the equation of state
of spin polarized SQM becomes stiffer.
Finally, using the general relativistic TOV equation, we studied the structure of the
spin polarized strange quark star (SQS). We calculated the maximum mass and the corresponding
radius of spin polarized SQS at different temperatures and magnetic fields.
We have shown that the gravitational mass
of spin polarized SQS rapidly increases by increasing the central energy density.
We also observed that the mass and the radius of the spin polarized SQS decreases by
increasing both temperature and magnetic field.
%%%%%%%%%%%%%%%%%%%%%%%%%%%%%%%%%%%%%%%%%%%%%%%%%%%%%%%%%%%%%%%%%%%%%%%%%%%%%%%%%%%%%%%%%%%%%%%%%%
\acknowledgements{ We wish to thank the Research
Institute for Astronomy and Astrophysics of Maragha and Shiraz University Research Council for financial support.}

%%%%%%%%%%%%%%%%%%%%%%%%%%%%%%%%%%%%%%%%%%%%%%%%%%%%%%%%%%%%%%%%%%%%%%%%%%%%%%%%%%%%%%%%%%%%%%%%%%

%%%%%%%%%%%%%%%%%%%%%%%%%%%%%%%%%%%%%%%%%%%%%%%%%%%%%%%%%%%%%%%%%%%%%%%%%%%%%%%%%%%%%%%%%%%%%%%%%%%%%%

%%%%%%%%%%%%%%%%%%%%%%%%%%%%%%%%%%%%%%%%%%%%%%%%%%%%%%%%%%%%%%%%%%%%%%%%%%%%%%%%%%%%%%%%
\newpage
\begin{table}[h]
\begin{center}
  \caption[]{Maximum mass and the corresponding radius of SQS for
  $B = 5 \times 10^{18}\ G$ at different temperatures.
 The results of  $T=0\ MeV$ (Bordbar \& Peyvand~\cite{rk8}) have been
  also given for comparison.}\label{T1}
  \begin{tabular}{clclclc}
  \hline\noalign{\smallskip}
 $T\ (MeV)$ & $M_{max}\ (M_{\odot})$ & $R\ (km)$  \\
 \hline\noalign{\smallskip}
 $0$ & 1.33 & 7.44   \\
 $30$ & 1.17 & 7.37 \\
 $70$ & 1.16 & 7.21   \\
  \noalign{\smallskip}\hline
  \end{tabular}
\end{center}
\end{table}
%=========================================================
\begin{table}[h]
\begin{center}
  \caption[]{Maximum mass and the corresponding radius of SQS for different
  magnetic fields at $T=30\ MeV$.}\label{T2}
  \begin{tabular}{clclcl}
  \hline\noalign{\smallskip}
 $B\ (G)$ & $M_{max}\ (M_{\odot})$ & $R\ (km)$  \\
 \hline\noalign{\smallskip}
 $0$ & 1.40 & 8.08 \\
 $5\times10^{18}$ & 1.17 & 7.37   \\
 $5\times10^{19}$ & 1.16 & 7.16 \\
 \noalign{\smallskip}\hline
  \end{tabular}
\end{center}
\end{table}
%%%%%%%%%%%%%%%%%%%%%%%%%%%%%%%%%%%%%%%%%%%%
\newpage
%%%%%%%%%%%%%%%%%%%%%%%%%%%%%%%%%%%%%%%%%%%%%%%%%%%%%%%%%%%%%%%%%%%%%%%%%%%%%%%%%%%%%%%%%%%%%%%%%%%%%%
\newpage
%%%%%%%%%%%%%%%%%%%%%%%%%%%%%%%%%%%%%%%%%%%%%%%%%%%%%%%%%%%%%%%%%%%%%%%%%%%%%%%%%%%%%%%%%%%%%%%%%%%%%%
\newpage
%%%%%%%%%%%%%%%%%%%%%%%%%%%%%%%%%%%%%%%%%%%%%%%%%%%%%%%%%%%%%%%%%%%%%%%%%%%%%%%%%%%%%%%%%%%
\newpage
%%%%%%%%%%%%%%%%%%%%%%%%%%%%%%%%%%%%%%%%%%%%%%%%%%%%%%%%%%%%%%%%%%%%%%%%%%%%%%%%%%%%%%%%%%%
\newpage
%%%%%%%%%%%%%%%%%%%%%%%%%%%%%%%%%%%%%%%%%%%%%%%%%%%%%%%%%%%%%%%%%%%%%%%%%%%%%%%%%%%%%%%%%%%
\newpage
%%%%%%%%%%%%%%%%%%%%%%%%%%%%%%%%%%%%%%%%%%%%%%%%%%%%%%%%%%%%%%%%%%%%%%%%%%%%%%%%%%%%%%%%%%%
\newpage
%%%%%%%%%%%%%%%%%%%%%%%%%%%%%%%%%%%%%%%%%%%%%%%%%%%%%%%%%%%%%%%%%%%%%%%%%%%%%%%%%%%%%%%%%%%
\newpage
%%%%%%%%%%%%%%%%%%%%%%%%%%%%%%%%%%%%%%%%%%%%%%%%%%%%%%%%%%%%%%%%%%%%%%%%%%%%%%%%%%%%%%%%%%%
\newpage
%%%%%%%%%%%%%%%%%%%%%%%%%%%%%%%%%%%%%%%%%%%%%%%%%%%%%%%%%%%%%%%%%%%%%%%%%%%%%%%%%%%%%%%%%%%

%%%%%%%%%%%%%%%%%%%%%%%%%%%%%%%%%%%%%%%%%%%%%%%%%%%%%%%%%%%%%%%%%%%%%%%%%%%%%%%%%%%%%%%%%%%
\end{document}